\documentclass[cup7b]{cupbook}

\begin{document}

\newtheorem{theorem}{Theorem}[chapter]
\newcommand{\blackboard}{\bf }

\pagenumbering{roman}

\pagenumbering{arabic}

\author[W.\ Taylor]{W. TAYLOR\\ MIT, Stanford University
\\ \\
{\rm SU-ITP-06/14 \hspace*{0.1in} MIT-CTP-3747 \hspace*{0.1in} hep-th/0605202}
}
\chapter{String field theory} 
\begin{center}
\end{center}

\begin{abstract}
This elementary introduction to string field theory highlights the
features and the limitations of this approach to quantum gravity as it
is currently understood.  String field theory is a formulation of
string theory as a field theory in space-time with an infinite number
of massive fields.  Although existing constructions of string field
theory require expanding around a fixed choice of space-time
background, the theory is in principle background-independent, in the
sense that different backgrounds can be realized as different field
configurations in the theory.  String field theory is the only string
formalism developed so far which, in principle, has the potential to
systematically address questions involving multiple asymptotically
distinct string backgrounds.  Thus, although it is not yet well
defined as a quantum theory, string field theory may eventually be
helpful for understanding questions related to cosmology in string
theory.
\end{abstract}

\section{Introduction}

In the early days of the subject, string theory was understood only as
a perturbative theory.  The theory arose from the study of S-matrices
and was conceived of as a new class of theory describing perturbative
interactions of massless particles including the gravitational quanta,
as well as an infinite family of massive particles associated with
excited string states.  In string theory, instead of the
one-dimensional world line of a pointlike particle tracing out a path
through space-time, a two-dimensional surface describes the trajectory
of an oscillating loop of string, which appears pointlike only to an
observer much larger than the string.

As the theory developed further, the need for a nonperturbative
description of the theory became clear.  The M(atrix) model of
M-theory, and the AdS/CFT correspondence, each of which is reviewed in
another chapter of this volume, are nonperturbative descriptions of
string theory in space-time backgrounds with fixed asymptotic forms.
These approaches to string theory give true nonperturbative
formulations of the theory, which fulfill in some sense one of the
primary theoretical goals of string theory: the formulation of a
nonperturbative theory of quantum gravity.

There are a number of questions, however, which cannot--even in
principle--be answered using perturbative methods or the
nonperturbative M(atrix) and AdS/CFT descriptions.  Recent
experimental evidence points strongly to the conclusion that the
space-time in which we live has a small but nonzero positive
cosmological constant.  None of the existing formulations of string
theory can be used to describe physics in such a space-time, however,
existing tools in string theory and field theory suggest that string
theory has a large number of metastable local minima with positive
cosmological constants.  The term ``string landscape'' (see, {\it
e.g.}, Susskind,  2003) is often used to describe the space of string
theory configurations which includes all these metastable local
minima.  We currently have no tools to rigorously define this space of
string theory configurations, however, or to understand the dynamics
of string theory in a cosmological context--a formalism capable of
describing the string landscape would presumably need to be a
background-independent formulation of the theory such as string field
theory.

The traditional perturbative approach to string theory involves
constructing a field theory on the two-dimensional string
``world-sheet'' $\Sigma$, which is mapped into the ``target''
space-time through a function $X: \Sigma \rightarrow$ space-time; this
function is locally described by a set of coordinates $X^\mu$.  The
theory on the world-sheet is quantized, and the excitations of the
resulting string become associated with massless and massive particles
moving in space-time.  The states of the string live in a Fock space
similar to the state space of a quantized simple harmonic oscillator.
The ground state of the string at momentum $p$, denoted $| p \rangle$,
is associated with a space-time scalar particle\footnote{Actually,
this ground state is associated with a scalar tachyon field describing
a particle with negative mass squared $m^2 < 0$.  The presence of such
a tachyon indicates that the vacuum around which the theory is being
expanded is unstable.  This tachyon is removed from the spectrum when
we consider supersymmetric string theory. } of momentum $p$.  There
are two kinds of raising operators acting on the single-string Fock
space, analogous to the raising operator $a^{\dagger}$ which adds a
unit of energy to a simple harmonic oscillator.  The operators
$\alpha^\mu_{-n} = (\alpha_n^\mu)^{\dagger}$ and $
\tilde{\alpha}^\mu_{-n} =(\tilde{\alpha}_n^\mu)^{\dagger}$ each add a
unit of excitation to the $n$th oscillation modes of the $\mu$
coordinate of the string.  There are two operators for each $n$
because there are two such oscillation modes, which can be thought of
as $\sin$ and $\cos$ modes or as right- and left- moving modes.  The
excited states of the string correspond to different particles in
space-time.  For example, the state
\begin{equation}
(\alpha_{-1}^\mu \tilde{\alpha}_{-1}^\nu+
\alpha_{-1}^\nu \tilde{\alpha}_{-1}^\mu) | p \rangle
\label{eq:graviton}
\end{equation}
corresponds to a symmetric spin 2 particle of momentum $p$.  These
states satisfy a physical state condition $p^2 = 0$, so that this
excitation state of the string can be associated with a quantum of the
gravitational field---a graviton.  Acting with more raising operators
on the string state produces a series of more and more highly excited strings
corresponding to a tower of massive particle states in space-time.  In
perturbative string theory,  interactions between the massless and
massive particles of the theory are computed by calculating
correlation functions on the string world-sheet using techniques of
two-dimensional conformal field theory.

The basic idea of string field theory is to reformulate string theory
in the target space-time, rather than on the world sheet, as an off-shell
theory of the infinite number of fields associated with the states in
the string Fock space.  The degrees of freedom in string field theory
are encoded in a ``string field'', which can be thought of in several
equivalent ways.  Conceptually, the simplest way to think of a string
field is as a functional $\Psi[X (\sigma)]$, which associates a
complex number with every possible configuration $X(\sigma)$ of a
one-dimensional string with coordinate $\sigma$.  This is the natural
generalization to a string of the standard quantum mechanical wave
function $\psi (x)$, which associates a complex number with every
possible position $x$ of a pointlike particle in space.
Mathematically, however, dealing directly with functionals like
$\Psi[X (\sigma)]$ is difficult and awkward.  In most cases it is more
convenient to use a Fock space representation of the the string
field.  Just as a wave function $\psi(x) \in{\cal L}^2 ({\bf R})$
for a single particle can be represented in a basis of harmonic
oscillator eigenstates $| n \rangle = \frac{1}{\sqrt{n!}}
(a^{\dagger})^n | 0 \rangle$ through $\psi (x) \rightarrow \sum_{n}c_n
| n \rangle$, the string field $\Psi[X (\sigma)]$ representing a
string moving in $D$ space-time dimensions
can be equivalently
represented in the string Fock space through
\begin{equation}
\Psi = \int d^Dp [\phi (p) | p \rangle +
g_{\mu \nu} (p) 
(\alpha_{-1}^\mu \tilde{\alpha}_{-1}^\nu+
\alpha_{-1}^\nu \tilde{\alpha}_{-1}^\mu)| p \rangle + \cdots]
\label{eq:closed-expansion}
\end{equation}
where the sum includes contributions from the infinite tower of
massive string states.  Because in this case the states carry a
continuously varying momentum, the coefficient of each state, which
was just a constant $c_n$ in the case of the harmonic oscillator,
becomes a field in space-time written in the Fourier representation.
Thus, we see that the string field contains within it an infinite
family of space-time fields, including the scalar field $\phi$, the
graviton field (metric) $g_{\mu \nu}$, and an infinite family of
massive fields.

String field theory is defined by giving an action functional ${\cal
L} (\Psi)$ depending on the string field.  When written in terms of
the individual component fields $\phi (x), g_{\mu \nu} (x), \ldots$,
this then gives a fairly conventional-looking action for a quantum
field theory, although the number of fields
is infinite and the interactions may contain higher derivatives and
appear nonlocal.  To be a consistent  description
of a known perturbative string theory, the action must be chosen
carefully so that the perturbative string field theory diagrams precisely
reproduce the string amplitudes computed from the perturbative string
theory.  This requirement puts a highly constraining algebraic
structure on the theory  (Zwiebach,  1993; Gaberdiel and Zwiebach,
1997a, 1997b).  Generally, it is necessary to
include an infinite series of terms in the action to meet this
requirement, although in the case of the bosonic open string Witten
has given an elegant formulation of string field theory which includes
only cubic interaction terms for the string field $\Psi$.  We will
describe this simplest and best-understood string field theory in the
next section.

Once a string field theory has been defined through an action, the
next question is whether it can be used as a tool to usefully compute
new results in string theory which extend beyond those accessible to
the perturbative formulation of the theory.  Although work on string
field theory began over 30 years ago, until 7 years ago there was no
clear example of a calculation in which string field theory gave
results which go beyond perturbation theory.  In 1999, however, Ashoke
Sen (1999)
made an insightful conjecture that two distinct open string
backgrounds, one with a space filling D-brane and one without, could
be explicitly realized as different solutions of the same open string
field theory.  Subsequent work on this conjecture has brought new
impetus to the study of string field theory, and has conclusively
demonstrated the nonperturbative background-independence of the
theory.  Despite these advances, however, there are still enormous
technical challenges for the theory.  The theory is not completely
well-defined even at the classical level, and a full definition of the
quantum theory seems very difficult.  Analytic calculations are
difficult and involve subtle issues of limits and divergences, and
numerical computations, while possible in many cases, are cumbersome
and often difficult to interpret.  Even for the simpler open string
field theory many conceptual challenges exist, and although there has
been recent progress on formulating closed string field theories,
using these theories to describe the landscape of string vacua is
still well beyond our technical capacity.

In the remainder of this paper we describe in some further detail the state of
knowledge in this subject.  In section 2 we give a somewhat more
explicit description of Witten's open bosonic string field theory;
we describe the recent work in which this theory was shown to describe
distinct string backgrounds, and we discuss some outstanding issues
for this theory.  In section 3 we review the state of the
art in closed string field theory.  Section 4 contains a summary of
successes and challenges for this formulation of string theory and
some speculation about possible future directions for this area
of research
 
\section{Open string field theory (OSFT)}

We now introduce the simplest covariant string field theory.
A very simple cubic form for the
off-shell open bosonic string field theory action was proposed by
Witten (1986).
In subsection \ref{sec:Witten} we briefly summarize the string field
theory described by this action.  In subsection \ref{sec:tachyon} we
review the recent work applying this theory to the study of Sen's
conjecture and discuss the progress which has been made.
For a more detailed review of this subject see Taylor \& Zwiebach (2001).
In subsection \ref{sec:issues} we discuss some problems and outstanding
issues for open string field theory.

It is useful to recall here the difference between open and closed
strings.  A closed string forms a one-dimensional loop.
Parameterizing the string by $\sigma \in[0, 1]$ we form a closed
string by identifying the endpoints $\sigma = 0, \sigma = 1$.  Because
fields on a closed string take periodic boundary conditions, there are
separate right- and left-moving modes.  This is what allows us to
construct a graviton state from a closed string as in
(\ref{eq:graviton}).  An open string, on the other hand, has Dirichlet
($X = 0$) or Neumann ($\partial_\sigma X = 0$) boundary conditions at
the endpoints, and therefore only has one set of oscillation modes,
which are associated with a single family of raising operators
$\alpha^\mu_{-n}$.  For the bosonic open string, the string field can
then be expanded as
\begin{equation}
\Psi =
\int d^{26}p \;
\left[ \varphi (p)\; | p \rangle + A_\mu (p) \; \alpha^\mu_{-1} | p
\rangle + \cdots \right]\,.
            \label{eq:field-expansion}
\end{equation}
The leading fields in this expansion are a space-time tachyon field
$\varphi (p)$ and a massless space-time vector field $A_ \mu (p)$.

\subsection{Witten's cubic OSFT action}
\label{sec:Witten}

The action proposed by Witten for the open bosonic string field theory
takes the simple  cubic form
\begin{equation}
S = -\frac{1}{2}\int \Psi \star Q \Psi -\frac{g}{3}  \int \Psi \star
\Psi \star \Psi\,.
\label{eq:SFT-action}
\end{equation}
In this action, $g$ is the (open) string coupling constant.  The field
$\Psi$ is the open string field.  Abstractly, this field can be
considered to take values in  an algebra ${\mathcal A}$.  
Associated with the algebra ${\mathcal A}$ there is a star product
\begin{equation}
\star:{\mathcal A} \otimes{\mathcal A} \rightarrow{\mathcal A}, \;\;\;\;\;
\end{equation}
The algebra ${\cal A}$ is graded, such that the open string field has
degree $G = 1$, and the degree $G$ is additive under the star product
($G_{\Psi \star \Phi} = G_\Psi + G_\Phi$).  There is also an
operator
\begin{equation}
Q:{\mathcal A} \rightarrow{\mathcal A}, \;\;\;\;\;
\end{equation}
called the BRST operator, which is
of degree one ($G_{Q \Psi} = 1 + G_\Psi$).  String fields can be
integrated using
\begin{equation}
\int:{\mathcal A} \rightarrow {\bf C}\,.
\end{equation}
This integral vanishes for all $\Psi$ with degree $G_\Psi \neq 3$.
Thus, the action (\ref{eq:SFT-action}) is only nonvanishing for a
string field $\Psi$ of degree 1.  The action (\ref{eq:SFT-action})
thus has the general form of a Chern-Simons theory on a 3-manifold,
although for string field theory there is no explicit interpretation
of the integration in terms of a concrete 3-manifold.

The elements $Q, \star, \int$ that define the string field theory are
assumed to satisfy the following axioms:
\vspace*{0.15in}

\noindent {\bf (a)} Nilpotency of $Q$: $\;Q^2 \Psi = 0, \;\; \; \forall \Psi
\in{\mathcal A}$.
\vspace*{0.08in}

\noindent {\bf (b)} $\int Q\Psi = 0, \; \; \; \forall \Psi \in{\mathcal A}$.
\vspace*{0.08in}

\noindent {\bf (c)} Derivation property of $Q$:\\
\hspace*{0.4in}$\;Q (\Psi \star \Phi) = (Q \Psi) \star \Phi +
(-1)^{G_\Psi} \Psi \star (Q \Phi), \; \; \forall \Psi, \Phi \in{\mathcal A}$.
\vspace*{0.08in}

\noindent {\bf (d)} Cyclicity:  $\;\int \Psi \star \Phi = (-1)^{G_\Psi
G_\Phi} \int \Phi \star \Psi, \; \; \; \forall \Psi, \Phi \in{\mathcal A}$.
\vspace*{0.08in}

\noindent {\bf (e)}  Associativity:  $(\Phi \star \Psi) \star \Xi =
\Phi \star (\Psi \star \Xi), \; \; \;
\forall \Phi, \Psi, \Xi \in{\mathcal A}$.
\vspace*{0.15in}

When these axioms are satisfied, the action (\ref{eq:SFT-action}) is
invariant under the gauge transformations
\begin{equation}
\delta \Psi = Q \Lambda + \Psi\star \Lambda - \Lambda \star \Psi\,,
            \label{eq:SFT-gauge}
\end{equation}
for any gauge parameter $\Lambda \in{\mathcal A}$ with degree 0.

When the string coupling $g$ is taken to vanish, the equation of
motion for the theory defined by (\ref{eq:SFT-action}) simply becomes
$Q \Psi = 0$, and the gauge transformations (\ref{eq:SFT-gauge})
simply become
\begin{equation}
\delta \Psi = Q \Lambda\,.
\end{equation}
This structure at $g = 0$ is precisely what is needed to describe a
free bosonic string in the BRST formalism, where physical states live
in the cohomology of the BRST operator $Q$, which acts on the string
Fock space\footnote{For a detailed introduction to BRST string
quantization, see Polchinski (1998)}.  The motivation for introducing
the extra structure in (\ref{eq:SFT-action}) was to find a simple
interacting extension of the free theory, consistent with the
perturbative expansion of open bosonic string theory.

Witten presented this formal structure and argued
that all the needed axioms are satisfied when ${\mathcal A}$ is taken to
be the space of string fields of the form (\ref{eq:field-expansion}).
In this realization, the star product $\star$ acts on a pair
of functionals $\Psi, \Phi$ by gluing the right half of one string to
the left half of the other using a delta function interaction

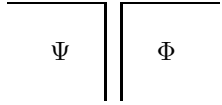
\begin{center}
\begin{picture}(100,60)(- 50,- 30)
\put(-40,20){\line(1,0){37}}
\put(40,20){\line(-1,0){37}}
\put(-3,20){\line( 0, -1){ 37}}
\put(3,20){\line( 0, -1){ 37}}
\put(-20,2){\makebox(0,0){$\Psi$}}
\put(20, 2){\makebox(0,0){$\Phi$}}
\end{picture}
\end{center}

Similarly, the integral over a string field corresponds to gluing the
left and right halves of the string together with a delta function
interaction

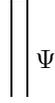
\begin{center}
\begin{picture}(100,60)(- 50,- 30)
\put(-3,20){\line(1,0){6}}
\put(-3,20){\line( 0, -1){ 37}}
\put(3,20){\line( 0, -1){ 37}}
\put(10,-3){\makebox(0,0){$\Psi$}}
\end{picture}
\end{center}

Combining these pictures,
the three-string vertex $\int \Psi_1 \star \Psi_2 \star \Psi_3$
corresponds to a three-string overlap

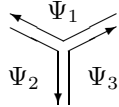
\begin{center}
\centering
\begin{picture}(100,60)(- 50,- 30)
\put( 20,14){\line( -2,-1){20}}
\put(0, 4){\vector( -2,1){20}}
\put(-20,9){\line( 2, -1){18}}
\put( -2, 0 ){\vector( 0, -1){ 20}}
\put( 2,-20){\line( 0, 1){ 20}}
\put(2,0){\vector( 2,1){18}}
\put(-15,-10){\makebox(0,0){$\Psi_2$}}
\put(0, 15){\makebox(0,0){$\Psi_1$}}
\put(15, -10){\makebox(0,0){$\Psi_3$}}
\end{picture}
\end{center}

While these pictures may seem rather abstract, they can be given
explicit meaning in terms of the oscillator raising and lowering
operators $\alpha^\mu_n$ (Cremmer
{\it et al.}, 1986; Ohta, 1986; Samuel, 1986;
Gross and Jevicki, 1987a, 1987b).  Given an
explicit representation of the terms in the string field action in
terms of these raising and lowering operators, the contribution to the
action from any set of component fields in the full string field can
be worked out.  The quadratic terms for the string fields $\varphi
(p), A_\mu (p)$ are the standard kinetic and mass terms for a tachyon
field and a massless gauge field.  The massive string fields similarly
have kinetic terms and positive mass squared terms.  The interaction
terms for the component fields coming from the term $ \int \Psi \star
\Psi \star \Psi$ in the action, however, seem more exotic from the
point of view of conventional field theory.  These terms contain
exponentials of derivatives, which appear as nonlocal interactions
from the point of view of field theory.  For example, the cubic
interaction term for the scalar tachyon field $\varphi (p)$ takes the
momentum space form
\begin{equation}
\int d^{26}pd^{26}q \; \frac{\kappa g}{3}  \;
           e^{(\ln 16/27) (p^2 + q^2 + p \cdot q)}
\varphi (-p) \varphi (-q) \varphi (p + q) \,.
\end{equation}
where $\kappa$ is a constant.
There are similar interaction terms between general sets of 3
component fields in the string field.

The appearance of an infinite number of fields and arbitrary numbers
of derivatives (powers of momentum) in the action make the target
space string field theory into a very unusual field theory. There are
a number of obstacles to having a complete definition of this theory
as a quantum field theory.  Even at the classical level, it is not
clear precisely what range of fields is allowed for the string field.
In particular, due to the presence of ghosts, there is no positive
definite inner product on the string Fock space, so there is no
natural finite norm condition to constrain the class of allowed string
fields.  Determining precisely what normalization condition should be
satisfied by physical states is an important problem which may need to
be solved to make substantial progress with the theory as a
nonperturbative formulation of string theory.  Beyond this
issue the unbounded number of derivatives makes even the classical
time-dependence of the string field difficult to pin down.  The string
field seems to obey a differential equation of infinite order,
suggesting an infinite number of boundary conditions are needed.  Some
recent progress on these problems has been made (Moeller \& Zwiebach,
2002; 
Erler \& Gross, 2004;
Coletti {\it {\it et al.}}, 2005), but
even in this simplest case of Witten's open cubic bosonic string field
theory, it seems clear that we are far from a complete understanding
of how the theory should be defined.
Despite these difficulties, however, the action (\ref{eq:SFT-action})
gives rise to a well-defined perturbative theory which can be used to
calculate scattering amplitudes of on-shell string states associated
with particles in the string Fock space.  Furthermore, it was shown
that these amplitudes agree with  the perturbative
formulation of 
string theory, as desired (Giddings \& Martinec, 1986; Giddings,
Martinec, \& Witten, 1986; 
Zwiebach, 1991).

\subsection{The Sen conjectures}
\label{sec:tachyon}

Despite our limited understanding of the full definition of quantum
string field theory, in the last few years a great deal of progress
has been made in understanding the nature of the classical open string
field theory described in the previous subsection.

One apparent problem for the open bosonic string and the associated
string field theory is the open string tachyon.  This tachyon
indicates that the vacuum of the theory is unstable and can decay.
Ashoke Sen (1999) conjectured that a precise understanding of the
nature of this instability and decay process could be attained through
open string field theory.  He argued that the unstable vacuum is one
with a space-filling ``D-brane'' carrying positive energy density.
D-branes have been a major subject of study in string theory over the
last decade.  D-branes are higher-dimensional extended objects on
which open strings can end.  In supersymmetric string theories,
D-branes of some dimensions can be stable and supersymmetric.  In the
bosonic string theory, however, all D-branes are unstable.  Sen
suggested that the instability of the space-filling D-brane in bosonic
string theory is manifested by the open bosonic string tachyon.  He
further suggested that string field theory should contain another
nonzero field configuration $\Psi_*$ which would satisfy the classical
equation of motion $Q \Psi_*+ g \Psi_*\star \Psi_*= 0$.  Sen argued
that this nontrivial vacuum field configuration should have several
specific properties.  It should have a vacuum energy which is lower
than the initial unstable vacuum by precisely the volume of space-time
times the energy density (tension) $T$ of the unstable D-brane.  The
stable vacuum should also have no open string excitations.  This
latter condition is highly nontrivial and states that at the
linearized level all open string fluctuations around the nontrivial
vacuum become unphysical.  To realize this change of backgrounds, the
degrees of freedom of the theory must reorganize completely in going
from one background to another.  The ability of a single set of
degrees of freedom to rearrange themselves to form the physical
degrees of freedom associated with fluctuations around different
backgrounds is perhaps the most striking feature of
background-independent theories, and presents the greatest challenge
in constructing and understanding such theories.

\begin{figure}
\setlength{\unitlength}{0.15pt}
\ifx\plotpoint\undefined\newsavebox{\plotpoint}\fi
\sbox{\plotpoint}{\rule[-0.200pt]{0.400pt}{0.400pt}}%
\begin{picture}(1500,1100)(80,0)
\font\gnuplot=cmr10 at 10pt
\gnuplot
\sbox{\plotpoint}{\rule[-0.200pt]{0.400pt}{0.400pt}}%
\put(100.0,675.0){\rule[-0.200pt]{4.818pt}{0.400pt}}
\put(80,675){\makebox(0,0)[r]{ 0}}
\put(1119.0,675.0){\rule[-0.200pt]{4.818pt}{0.400pt}}
\put(467.0,82.0){\rule[-0.200pt]{0.400pt}{4.818pt}}
\put(467,41){\makebox(0,0){ 0}}
\put(467.0,740.0){\rule[-0.200pt]{0.400pt}{4.818pt}}
\put(100.0,675.0){\rule[-0.200pt]{222.565pt}{0.400pt}}
\put(1384,638){\makebox(0,0)[l]{$\varphi$}}
\put(325,941){\makebox(0,0)[l]{$V (\varphi)$}}
\put(467.0,82.0){\rule[-0.200pt]{0.400pt}{147.420pt}}
\sbox{\plotpoint}{\rule[-0.400pt]{0.800pt}{0.800pt}}%
\put(1379,920){\makebox(0,0)[r]{Effective tachyon potential}}
\put(1399.0,920.0){\rule[-0.400pt]{24.090pt}{0.800pt}}
\put(238,477){\usebox{\plotpoint}}
\multiput(239.40,477.00)(0.516,1.179){11}{\rule{0.124pt}{1.978pt}}
\multiput(236.34,477.00)(9.000,15.895){2}{\rule{0.800pt}{0.989pt}}
\multiput(248.40,497.00)(0.516,0.990){11}{\rule{0.124pt}{1.711pt}}
\multiput(245.34,497.00)(9.000,13.449){2}{\rule{0.800pt}{0.856pt}}
\multiput(257.40,514.00)(0.516,0.927){11}{\rule{0.124pt}{1.622pt}}
\multiput(254.34,514.00)(9.000,12.633){2}{\rule{0.800pt}{0.811pt}}
\multiput(266.40,530.00)(0.516,0.863){11}{\rule{0.124pt}{1.533pt}}
\multiput(263.34,530.00)(9.000,11.817){2}{\rule{0.800pt}{0.767pt}}
\multiput(275.40,545.00)(0.516,0.800){11}{\rule{0.124pt}{1.444pt}}
\multiput(272.34,545.00)(9.000,11.002){2}{\rule{0.800pt}{0.722pt}}
\multiput(284.40,559.00)(0.514,0.654){13}{\rule{0.124pt}{1.240pt}}
\multiput(281.34,559.00)(10.000,10.426){2}{\rule{0.800pt}{0.620pt}}
\multiput(294.40,572.00)(0.516,0.674){11}{\rule{0.124pt}{1.267pt}}
\multiput(291.34,572.00)(9.000,9.371){2}{\rule{0.800pt}{0.633pt}}
\multiput(303.40,584.00)(0.516,0.548){11}{\rule{0.124pt}{1.089pt}}
\multiput(300.34,584.00)(9.000,7.740){2}{\rule{0.800pt}{0.544pt}}
\multiput(312.40,594.00)(0.516,0.548){11}{\rule{0.124pt}{1.089pt}}
\multiput(309.34,594.00)(9.000,7.740){2}{\rule{0.800pt}{0.544pt}}
\multiput(321.40,604.00)(0.516,0.548){11}{\rule{0.124pt}{1.089pt}}
\multiput(318.34,604.00)(9.000,7.740){2}{\rule{0.800pt}{0.544pt}}
\multiput(329.00,615.40)(0.554,0.520){9}{\rule{1.100pt}{0.125pt}}
\multiput(329.00,612.34)(6.717,8.000){2}{\rule{0.550pt}{0.800pt}}
\multiput(338.00,623.40)(0.627,0.520){9}{\rule{1.200pt}{0.125pt}}
\multiput(338.00,620.34)(7.509,8.000){2}{\rule{0.600pt}{0.800pt}}
\multiput(348.00,631.40)(0.650,0.526){7}{\rule{1.229pt}{0.127pt}}
\multiput(348.00,628.34)(6.450,7.000){2}{\rule{0.614pt}{0.800pt}}
\multiput(357.00,638.39)(0.797,0.536){5}{\rule{1.400pt}{0.129pt}}
\multiput(357.00,635.34)(6.094,6.000){2}{\rule{0.700pt}{0.800pt}}
\multiput(366.00,644.39)(0.797,0.536){5}{\rule{1.400pt}{0.129pt}}
\multiput(366.00,641.34)(6.094,6.000){2}{\rule{0.700pt}{0.800pt}}
\multiput(375.00,650.38)(1.096,0.560){3}{\rule{1.640pt}{0.135pt}}
\multiput(375.00,647.34)(5.596,5.000){2}{\rule{0.820pt}{0.800pt}}
\multiput(384.00,655.38)(1.096,0.560){3}{\rule{1.640pt}{0.135pt}}
\multiput(384.00,652.34)(5.596,5.000){2}{\rule{0.820pt}{0.800pt}}
\put(393,659.34){\rule{2.200pt}{0.800pt}}
\multiput(393.00,657.34)(5.434,4.000){2}{\rule{1.100pt}{0.800pt}}
\put(403,662.84){\rule{2.168pt}{0.800pt}}
\multiput(403.00,661.34)(4.500,3.000){2}{\rule{1.084pt}{0.800pt}}
\put(412,665.84){\rule{2.168pt}{0.800pt}}
\multiput(412.00,664.34)(4.500,3.000){2}{\rule{1.084pt}{0.800pt}}
\put(421,668.34){\rule{2.168pt}{0.800pt}}
\multiput(421.00,667.34)(4.500,2.000){2}{\rule{1.084pt}{0.800pt}}
\put(430,670.34){\rule{2.168pt}{0.800pt}}
\multiput(430.00,669.34)(4.500,2.000){2}{\rule{1.084pt}{0.800pt}}
\put(439,671.84){\rule{2.409pt}{0.800pt}}
\multiput(439.00,671.34)(5.000,1.000){2}{\rule{1.204pt}{0.800pt}}
\put(449,672.84){\rule{2.168pt}{0.800pt}}
\multiput(449.00,672.34)(4.500,1.000){2}{\rule{1.084pt}{0.800pt}}
\put(476,672.84){\rule{2.168pt}{0.800pt}}
\multiput(476.00,673.34)(4.500,-1.000){2}{\rule{1.084pt}{0.800pt}}
\put(485,671.84){\rule{2.168pt}{0.800pt}}
\multiput(485.00,672.34)(4.500,-1.000){2}{\rule{1.084pt}{0.800pt}}
\put(494,670.34){\rule{2.409pt}{0.800pt}}
\multiput(494.00,671.34)(5.000,-2.000){2}{\rule{1.204pt}{0.800pt}}
\put(504,668.34){\rule{2.168pt}{0.800pt}}
\multiput(504.00,669.34)(4.500,-2.000){2}{\rule{1.084pt}{0.800pt}}
\put(513,666.34){\rule{2.168pt}{0.800pt}}
\multiput(513.00,667.34)(4.500,-2.000){2}{\rule{1.084pt}{0.800pt}}
\put(522,663.84){\rule{2.168pt}{0.800pt}}
\multiput(522.00,665.34)(4.500,-3.000){2}{\rule{1.084pt}{0.800pt}}
\put(531,660.84){\rule{2.168pt}{0.800pt}}
\multiput(531.00,662.34)(4.500,-3.000){2}{\rule{1.084pt}{0.800pt}}
\put(540,657.84){\rule{2.168pt}{0.800pt}}
\multiput(540.00,659.34)(4.500,-3.000){2}{\rule{1.084pt}{0.800pt}}
\put(549,654.34){\rule{2.200pt}{0.800pt}}
\multiput(549.00,656.34)(5.434,-4.000){2}{\rule{1.100pt}{0.800pt}}
\put(559,650.34){\rule{2.000pt}{0.800pt}}
\multiput(559.00,652.34)(4.849,-4.000){2}{\rule{1.000pt}{0.800pt}}
\multiput(568.00,648.06)(1.096,-0.560){3}{\rule{1.640pt}{0.135pt}}
\multiput(568.00,648.34)(5.596,-5.000){2}{\rule{0.820pt}{0.800pt}}
\multiput(577.00,643.06)(1.096,-0.560){3}{\rule{1.640pt}{0.135pt}}
\multiput(577.00,643.34)(5.596,-5.000){2}{\rule{0.820pt}{0.800pt}}
\multiput(586.00,638.06)(1.096,-0.560){3}{\rule{1.640pt}{0.135pt}}
\multiput(586.00,638.34)(5.596,-5.000){2}{\rule{0.820pt}{0.800pt}}
\multiput(595.00,633.06)(1.096,-0.560){3}{\rule{1.640pt}{0.135pt}}
\multiput(595.00,633.34)(5.596,-5.000){2}{\rule{0.820pt}{0.800pt}}
\multiput(604.00,628.07)(0.909,-0.536){5}{\rule{1.533pt}{0.129pt}}
\multiput(604.00,628.34)(6.817,-6.000){2}{\rule{0.767pt}{0.800pt}}
\multiput(614.00,622.07)(0.797,-0.536){5}{\rule{1.400pt}{0.129pt}}
\multiput(614.00,622.34)(6.094,-6.000){2}{\rule{0.700pt}{0.800pt}}
\multiput(623.00,616.07)(0.797,-0.536){5}{\rule{1.400pt}{0.129pt}}
\multiput(623.00,616.34)(6.094,-6.000){2}{\rule{0.700pt}{0.800pt}}
\multiput(632.00,610.07)(0.797,-0.536){5}{\rule{1.400pt}{0.129pt}}
\multiput(632.00,610.34)(6.094,-6.000){2}{\rule{0.700pt}{0.800pt}}
\multiput(641.00,604.08)(0.650,-0.526){7}{\rule{1.229pt}{0.127pt}}
\multiput(641.00,604.34)(6.450,-7.000){2}{\rule{0.614pt}{0.800pt}}
\multiput(650.00,597.08)(0.650,-0.526){7}{\rule{1.229pt}{0.127pt}}
\multiput(650.00,597.34)(6.450,-7.000){2}{\rule{0.614pt}{0.800pt}}
\multiput(659.00,590.08)(0.738,-0.526){7}{\rule{1.343pt}{0.127pt}}
\multiput(659.00,590.34)(7.213,-7.000){2}{\rule{0.671pt}{0.800pt}}
\multiput(669.00,583.08)(0.650,-0.526){7}{\rule{1.229pt}{0.127pt}}
\multiput(669.00,583.34)(6.450,-7.000){2}{\rule{0.614pt}{0.800pt}}
\multiput(678.00,576.08)(0.554,-0.520){9}{\rule{1.100pt}{0.125pt}}
\multiput(678.00,576.34)(6.717,-8.000){2}{\rule{0.550pt}{0.800pt}}
\multiput(687.00,568.08)(0.650,-0.526){7}{\rule{1.229pt}{0.127pt}}
\multiput(687.00,568.34)(6.450,-7.000){2}{\rule{0.614pt}{0.800pt}}
\multiput(696.00,561.08)(0.554,-0.520){9}{\rule{1.100pt}{0.125pt}}
\multiput(696.00,561.34)(6.717,-8.000){2}{\rule{0.550pt}{0.800pt}}
\multiput(705.00,553.08)(0.554,-0.520){9}{\rule{1.100pt}{0.125pt}}
\multiput(705.00,553.34)(6.717,-8.000){2}{\rule{0.550pt}{0.800pt}}
\multiput(714.00,545.08)(0.627,-0.520){9}{\rule{1.200pt}{0.125pt}}
\multiput(714.00,545.34)(7.509,-8.000){2}{\rule{0.600pt}{0.800pt}}
\multiput(724.00,537.08)(0.485,-0.516){11}{\rule{1.000pt}{0.124pt}}
\multiput(724.00,537.34)(6.924,-9.000){2}{\rule{0.500pt}{0.800pt}}
\multiput(733.00,528.08)(0.554,-0.520){9}{\rule{1.100pt}{0.125pt}}
\multiput(733.00,528.34)(6.717,-8.000){2}{\rule{0.550pt}{0.800pt}}
\multiput(742.00,520.08)(0.554,-0.520){9}{\rule{1.100pt}{0.125pt}}
\multiput(742.00,520.34)(6.717,-8.000){2}{\rule{0.550pt}{0.800pt}}
\multiput(751.00,512.08)(0.485,-0.516){11}{\rule{1.000pt}{0.124pt}}
\multiput(751.00,512.34)(6.924,-9.000){2}{\rule{0.500pt}{0.800pt}}
\multiput(760.00,503.08)(0.485,-0.516){11}{\rule{1.000pt}{0.124pt}}
\multiput(760.00,503.34)(6.924,-9.000){2}{\rule{0.500pt}{0.800pt}}
\multiput(769.00,494.08)(0.627,-0.520){9}{\rule{1.200pt}{0.125pt}}
\multiput(769.00,494.34)(7.509,-8.000){2}{\rule{0.600pt}{0.800pt}}
\multiput(779.00,486.08)(0.485,-0.516){11}{\rule{1.000pt}{0.124pt}}
\multiput(779.00,486.34)(6.924,-9.000){2}{\rule{0.500pt}{0.800pt}}
\multiput(788.00,477.08)(0.485,-0.516){11}{\rule{1.000pt}{0.124pt}}
\multiput(788.00,477.34)(6.924,-9.000){2}{\rule{0.500pt}{0.800pt}}
\multiput(797.00,468.08)(0.485,-0.516){11}{\rule{1.000pt}{0.124pt}}
\multiput(797.00,468.34)(6.924,-9.000){2}{\rule{0.500pt}{0.800pt}}
\multiput(806.00,459.08)(0.485,-0.516){11}{\rule{1.000pt}{0.124pt}}
\multiput(806.00,459.34)(6.924,-9.000){2}{\rule{0.500pt}{0.800pt}}
\multiput(815.00,450.08)(0.548,-0.516){11}{\rule{1.089pt}{0.124pt}}
\multiput(815.00,450.34)(7.740,-9.000){2}{\rule{0.544pt}{0.800pt}}
\multiput(825.00,441.08)(0.485,-0.516){11}{\rule{1.000pt}{0.124pt}}
\multiput(825.00,441.34)(6.924,-9.000){2}{\rule{0.500pt}{0.800pt}}
\multiput(834.00,432.08)(0.485,-0.516){11}{\rule{1.000pt}{0.124pt}}
\multiput(834.00,432.34)(6.924,-9.000){2}{\rule{0.500pt}{0.800pt}}
\multiput(843.00,423.08)(0.485,-0.516){11}{\rule{1.000pt}{0.124pt}}
\multiput(843.00,423.34)(6.924,-9.000){2}{\rule{0.500pt}{0.800pt}}
\multiput(852.00,414.08)(0.485,-0.516){11}{\rule{1.000pt}{0.124pt}}
\multiput(852.00,414.34)(6.924,-9.000){2}{\rule{0.500pt}{0.800pt}}
\multiput(861.00,405.08)(0.485,-0.516){11}{\rule{1.000pt}{0.124pt}}
\multiput(861.00,405.34)(6.924,-9.000){2}{\rule{0.500pt}{0.800pt}}
\multiput(870.00,396.08)(0.548,-0.516){11}{\rule{1.089pt}{0.124pt}}
\multiput(870.00,396.34)(7.740,-9.000){2}{\rule{0.544pt}{0.800pt}}
\multiput(880.00,387.08)(0.485,-0.516){11}{\rule{1.000pt}{0.124pt}}
\multiput(880.00,387.34)(6.924,-9.000){2}{\rule{0.500pt}{0.800pt}}
\multiput(889.00,378.08)(0.485,-0.516){11}{\rule{1.000pt}{0.124pt}}
\multiput(889.00,378.34)(6.924,-9.000){2}{\rule{0.500pt}{0.800pt}}
\multiput(898.00,369.08)(0.554,-0.520){9}{\rule{1.100pt}{0.125pt}}
\multiput(898.00,369.34)(6.717,-8.000){2}{\rule{0.550pt}{0.800pt}}
\multiput(907.00,361.08)(0.485,-0.516){11}{\rule{1.000pt}{0.124pt}}
\multiput(907.00,361.34)(6.924,-9.000){2}{\rule{0.500pt}{0.800pt}}
\multiput(916.00,352.08)(0.485,-0.516){11}{\rule{1.000pt}{0.124pt}}
\multiput(916.00,352.34)(6.924,-9.000){2}{\rule{0.500pt}{0.800pt}}
\multiput(925.00,343.08)(0.627,-0.520){9}{\rule{1.200pt}{0.125pt}}
\multiput(925.00,343.34)(7.509,-8.000){2}{\rule{0.600pt}{0.800pt}}
\multiput(935.00,335.08)(0.554,-0.520){9}{\rule{1.100pt}{0.125pt}}
\multiput(935.00,335.34)(6.717,-8.000){2}{\rule{0.550pt}{0.800pt}}
\multiput(944.00,327.08)(0.485,-0.516){11}{\rule{1.000pt}{0.124pt}}
\multiput(944.00,327.34)(6.924,-9.000){2}{\rule{0.500pt}{0.800pt}}
\multiput(953.00,318.08)(0.554,-0.520){9}{\rule{1.100pt}{0.125pt}}
\multiput(953.00,318.34)(6.717,-8.000){2}{\rule{0.550pt}{0.800pt}}
\multiput(962.00,310.08)(0.650,-0.526){7}{\rule{1.229pt}{0.127pt}}
\multiput(962.00,310.34)(6.450,-7.000){2}{\rule{0.614pt}{0.800pt}}
\multiput(971.00,303.08)(0.554,-0.520){9}{\rule{1.100pt}{0.125pt}}
\multiput(971.00,303.34)(6.717,-8.000){2}{\rule{0.550pt}{0.800pt}}
\multiput(980.00,295.08)(0.627,-0.520){9}{\rule{1.200pt}{0.125pt}}
\multiput(980.00,295.34)(7.509,-8.000){2}{\rule{0.600pt}{0.800pt}}
\multiput(990.00,287.08)(0.650,-0.526){7}{\rule{1.229pt}{0.127pt}}
\multiput(990.00,287.34)(6.450,-7.000){2}{\rule{0.614pt}{0.800pt}}
\multiput(999.00,280.08)(0.650,-0.526){7}{\rule{1.229pt}{0.127pt}}
\multiput(999.00,280.34)(6.450,-7.000){2}{\rule{0.614pt}{0.800pt}}
\multiput(1008.00,273.08)(0.650,-0.526){7}{\rule{1.229pt}{0.127pt}}
\multiput(1008.00,273.34)(6.450,-7.000){2}{\rule{0.614pt}{0.800pt}}
\multiput(1017.00,266.08)(0.650,-0.526){7}{\rule{1.229pt}{0.127pt}}
\multiput(1017.00,266.34)(6.450,-7.000){2}{\rule{0.614pt}{0.800pt}}
\multiput(1026.00,259.07)(0.797,-0.536){5}{\rule{1.400pt}{0.129pt}}
\multiput(1026.00,259.34)(6.094,-6.000){2}{\rule{0.700pt}{0.800pt}}
\multiput(1035.00,253.07)(0.909,-0.536){5}{\rule{1.533pt}{0.129pt}}
\multiput(1035.00,253.34)(6.817,-6.000){2}{\rule{0.767pt}{0.800pt}}
\multiput(1045.00,247.07)(0.797,-0.536){5}{\rule{1.400pt}{0.129pt}}
\multiput(1045.00,247.34)(6.094,-6.000){2}{\rule{0.700pt}{0.800pt}}
\multiput(1054.00,241.07)(0.797,-0.536){5}{\rule{1.400pt}{0.129pt}}
\multiput(1054.00,241.34)(6.094,-6.000){2}{\rule{0.700pt}{0.800pt}}
\multiput(1063.00,235.06)(1.096,-0.560){3}{\rule{1.640pt}{0.135pt}}
\multiput(1063.00,235.34)(5.596,-5.000){2}{\rule{0.820pt}{0.800pt}}
\multiput(1072.00,230.06)(1.096,-0.560){3}{\rule{1.640pt}{0.135pt}}
\multiput(1072.00,230.34)(5.596,-5.000){2}{\rule{0.820pt}{0.800pt}}
\multiput(1081.00,225.06)(1.096,-0.560){3}{\rule{1.640pt}{0.135pt}}
\multiput(1081.00,225.34)(5.596,-5.000){2}{\rule{0.820pt}{0.800pt}}
\put(1090,218.34){\rule{2.200pt}{0.800pt}}
\multiput(1090.00,220.34)(5.434,-4.000){2}{\rule{1.100pt}{0.800pt}}
\put(1100,214.34){\rule{2.000pt}{0.800pt}}
\multiput(1100.00,216.34)(4.849,-4.000){2}{\rule{1.000pt}{0.800pt}}
\put(1109,210.34){\rule{2.000pt}{0.800pt}}
\multiput(1109.00,212.34)(4.849,-4.000){2}{\rule{1.000pt}{0.800pt}}
\put(1118,206.34){\rule{2.000pt}{0.800pt}}
\multiput(1118.00,208.34)(4.849,-4.000){2}{\rule{1.000pt}{0.800pt}}
\put(1127,202.84){\rule{2.168pt}{0.800pt}}
\multiput(1127.00,204.34)(4.500,-3.000){2}{\rule{1.084pt}{0.800pt}}
\put(1136,200.34){\rule{2.409pt}{0.800pt}}
\multiput(1136.00,201.34)(5.000,-2.000){2}{\rule{1.204pt}{0.800pt}}
\put(1146,198.34){\rule{2.168pt}{0.800pt}}
\multiput(1146.00,199.34)(4.500,-2.000){2}{\rule{1.084pt}{0.800pt}}
\put(1155,196.34){\rule{2.168pt}{0.800pt}}
\multiput(1155.00,197.34)(4.500,-2.000){2}{\rule{1.084pt}{0.800pt}}
\put(1164,194.84){\rule{2.168pt}{0.800pt}}
\multiput(1164.00,195.34)(4.500,-1.000){2}{\rule{1.084pt}{0.800pt}}
\put(1173,193.84){\rule{2.168pt}{0.800pt}}
\multiput(1173.00,194.34)(4.500,-1.000){2}{\rule{1.084pt}{0.800pt}}
\put(1182,192.84){\rule{2.168pt}{0.800pt}}
\multiput(1182.00,193.34)(4.500,-1.000){2}{\rule{1.084pt}{0.800pt}}
\put(458.0,675.0){\rule[-0.400pt]{4.336pt}{0.800pt}}
\put(1201,192.84){\rule{2.168pt}{0.800pt}}
\multiput(1201.00,192.34)(4.500,1.000){2}{\rule{1.084pt}{0.800pt}}
\put(1210,193.84){\rule{2.168pt}{0.800pt}}
\multiput(1210.00,193.34)(4.500,1.000){2}{\rule{1.084pt}{0.800pt}}
\put(1219,194.84){\rule{2.168pt}{0.800pt}}
\multiput(1219.00,194.34)(4.500,1.000){2}{\rule{1.084pt}{0.800pt}}
\put(1228,196.34){\rule{2.168pt}{0.800pt}}
\multiput(1228.00,195.34)(4.500,2.000){2}{\rule{1.084pt}{0.800pt}}
\put(1237,198.34){\rule{2.168pt}{0.800pt}}
\multiput(1237.00,197.34)(4.500,2.000){2}{\rule{1.084pt}{0.800pt}}
\put(1246,200.84){\rule{2.409pt}{0.800pt}}
\multiput(1246.00,199.34)(5.000,3.000){2}{\rule{1.204pt}{0.800pt}}
\put(1256,204.34){\rule{2.000pt}{0.800pt}}
\multiput(1256.00,202.34)(4.849,4.000){2}{\rule{1.000pt}{0.800pt}}
\put(1265,208.34){\rule{2.000pt}{0.800pt}}
\multiput(1265.00,206.34)(4.849,4.000){2}{\rule{1.000pt}{0.800pt}}
\multiput(1274.00,213.38)(1.096,0.560){3}{\rule{1.640pt}{0.135pt}}
\multiput(1274.00,210.34)(5.596,5.000){2}{\rule{0.820pt}{0.800pt}}
\multiput(1283.00,218.38)(1.096,0.560){3}{\rule{1.640pt}{0.135pt}}
\multiput(1283.00,215.34)(5.596,5.000){2}{\rule{0.820pt}{0.800pt}}
\multiput(1292.00,223.39)(0.797,0.536){5}{\rule{1.400pt}{0.129pt}}
\multiput(1292.00,220.34)(6.094,6.000){2}{\rule{0.700pt}{0.800pt}}
\multiput(1301.00,229.40)(0.738,0.526){7}{\rule{1.343pt}{0.127pt}}
\multiput(1301.00,226.34)(7.213,7.000){2}{\rule{0.671pt}{0.800pt}}
\multiput(1311.00,236.40)(0.650,0.526){7}{\rule{1.229pt}{0.127pt}}
\multiput(1311.00,233.34)(6.450,7.000){2}{\rule{0.614pt}{0.800pt}}
\multiput(1320.00,243.40)(0.554,0.520){9}{\rule{1.100pt}{0.125pt}}
\multiput(1320.00,240.34)(6.717,8.000){2}{\rule{0.550pt}{0.800pt}}
\multiput(1329.00,251.40)(0.485,0.516){11}{\rule{1.000pt}{0.124pt}}
\multiput(1329.00,248.34)(6.924,9.000){2}{\rule{0.500pt}{0.800pt}}
\multiput(1338.00,260.40)(0.485,0.516){11}{\rule{1.000pt}{0.124pt}}
\multiput(1338.00,257.34)(6.924,9.000){2}{\rule{0.500pt}{0.800pt}}
\multiput(1348.40,268.00)(0.516,0.548){11}{\rule{0.124pt}{1.089pt}}
\multiput(1345.34,268.00)(9.000,7.740){2}{\rule{0.800pt}{0.544pt}}
\multiput(1356.00,279.40)(0.487,0.514){13}{\rule{1.000pt}{0.124pt}}
\multiput(1356.00,276.34)(7.924,10.000){2}{\rule{0.500pt}{0.800pt}}
\multiput(1367.40,288.00)(0.516,0.674){11}{\rule{0.124pt}{1.267pt}}
\multiput(1364.34,288.00)(9.000,9.371){2}{\rule{0.800pt}{0.633pt}}
\multiput(1376.40,300.00)(0.516,0.674){11}{\rule{0.124pt}{1.267pt}}
\multiput(1373.34,300.00)(9.000,9.371){2}{\rule{0.800pt}{0.633pt}}
\multiput(1385.41,312.00)(0.506,0.728){29}{\rule{0.122pt}{1.356pt}}
\multiput(1382.34,312.00)(18.000,23.186){2}{\rule{0.800pt}{0.678pt}}
\multiput(1403.40,338.00)(0.516,0.863){11}{\rule{0.124pt}{1.533pt}}
\multiput(1400.34,338.00)(9.000,11.817){2}{\rule{0.800pt}{0.767pt}}
\multiput(1412.40,353.00)(0.514,0.766){13}{\rule{0.124pt}{1.400pt}}
\multiput(1409.34,353.00)(10.000,12.094){2}{\rule{0.800pt}{0.700pt}}
\multiput(1422.40,368.00)(0.516,0.927){11}{\rule{0.124pt}{1.622pt}}
\multiput(1419.34,368.00)(9.000,12.633){2}{\rule{0.800pt}{0.811pt}}
\multiput(1431.40,384.00)(0.516,0.990){11}{\rule{0.124pt}{1.711pt}}
\multiput(1428.34,384.00)(9.000,13.449){2}{\rule{0.800pt}{0.856pt}}
\put(1191.0,194.0){\rule[-0.400pt]{2.409pt}{0.800pt}}
\put(1439,401){\usebox{\plotpoint}}
\end{picture}
 \caption{The effective tachyon potential in open string field theory}
 \label{f:potential}
\end{figure}
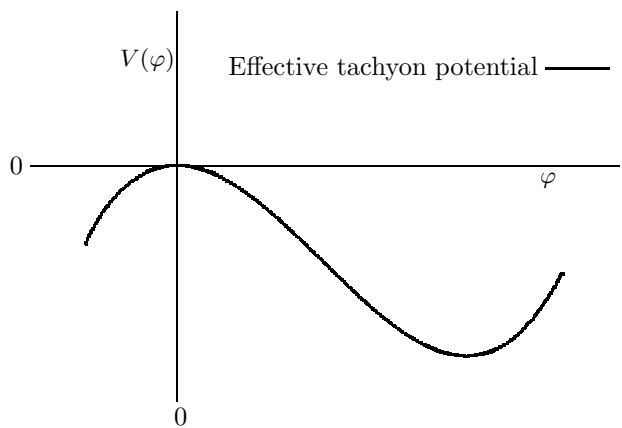

Following Sen's conjectures, a substantial body of work was carried
out which confirmed these conjectures in detail.  A primary tool used
in analyzing these conjectures using string field theory was the
notion of ``level truncation''.  The idea of level truncation is to
reduce the infinite number of string fields to a finite number by
throwing out all fields above a fixed mass cutoff.  By performing such
a truncation and restricting attention to the constant modes with $p =
0$, the infinite number of string field component equations reduces to
a finite system of cubic equations.  These equations were solved
numerically at various levels of truncation, and confirmed to
$99.99\%$ accuracy the conjecture that there is a nontrivial vacuum
solution with the predicted energy (Sen \& Zwiebach, 2000; Moeller \&
Taylor, 2000; Gaiotto \& Rastelli, 2003; Taylor, 2003).  The
conjecture that the nontrivial vacuum has no physical open string
excitations was also tested numerically and found to hold to high
accuracy (Ellwood \& Taylor, 2001a; Ellwood {\it et al.}, 2001).  The
effective potential $V (\varphi)$ for the tachyon field can be
computed using this approach;  this potential is graphed in
Figure~\ref{f:potential}.  This figure clearly illustrates the
unstable perturbative vacuum as well as the stable nonperturbative
vacuum.

\vspace*{0.1in}

The results of numerical analysis have confirmed Sen's conjectures
very clearly.  Perhaps the most important consequence of this
confirmation is that we have for the first time concrete evidence that
string field theory can describe multiple disconnected\footnote{By
  disconnected we mean that there is no continuous family of vacuum
  solutions interpolating between the distinct vacua.}
 string vacua in
terms of a common set of variables.  This is in principle the kind of
construction which is needed to describe the disparate string vacua of
the closed string landscape.  Indeed, Figure~\ref{f:potential} can be
seen as a piece of the ``open string landscape''.  To extrapolate from
the results achieved so far in classical open string field theory to
the picture we desire of a set of independent solutions of a quantum
closed string field theory, however, a number of significant further
steps must be taken.  We discuss some of the issues which must be
resolved in the following subsection.

\subsection{Outstanding problems and issues in OSFT}
\label{sec:issues}

In order to improve our understanding of OSFT so that we can better
understand the space of solutions of the theory, one very important
first step is to develop analytic tools to describe the nontrivial
open string vacuum described in the previous subsection.  One approach
to this problem was to try to reformulate string field theory around
this vacuum using ``vacuum string field theory'' (Rastelli {\it et
al.}, 2001).  This approach led to the development of some powerful
analytic tools for understanding the star algebra and projectors in
the theory; recently these tools were used to make an important step
forward by Schnabl (2005), who has found an analytic form for the
nontrivial vacuum of Witten's open string field theory.  The
presentation of this vacuum state has interesting analytic properties
related to Bernoulli numbers.  It seems to have a part which is
well-behaved under level truncation, and another part which involves
an infinite sequence of massive string fields.  The second part of
this state has vanishing inner product with all states which appear in
level truncation, and is not yet completely understood (for further
discussion of this state see Okawa, 2006).  This construction seems to
be a promising step towards developing analytic machinery to describe
solutions of classical string field theory; it seems likely that in
the reasonably near future this may lead to significant new
developments in this area.

Another important issue, relevant for understanding string field theory
analytically and for describing a disparate family of solutions to the
theory, even at the classical level, is the problem of field redefinitions.
The issue here is that the fields appearing in the string field, such
as $\varphi$ and $A_\mu$, are only identified at linear order with the
usual space-time fields of conformal field theory.  At higher order,
these fields are related by a highly nontrivial field redefinition
which can include arbitrary numbers of derivatives (Ghoshal \& Sen, 1992).
For example, the SFT $A_{\mu}$ (after integrating out the massive
fields) is related to the CFT $\tilde{A}_\mu$ by a field redefinition
\begin{equation}
\tilde{A}_\mu =  A_\mu + \alpha A^2 A_\mu + \beta A^2 \partial^2 A_\mu
+ \cdots
\end{equation}
where arbitrarily complicated terms appear on the RHS (Coletti {\it et
al.}, 2003).  Because of these field redefinitions, simple physical
properties such as turning on a constant deformation $A_\mu$,
corresponding to the simple translation of a D-brane in flat space in
a dual picture, are difficult to understand in the variables natural
to SFT (Sen \& Zwiebach, 2000).  Similar field redefinitions,
involving arbitrary numbers of time derivatives, take a reasonably
well-behaved time-dependent tachyon solution which classically rolls
down the hill depicted in Figure~\ref{f:potential} in the CFT
description to a string field theory solution which has wild
exponentially increasing oscillations (Moeller \& Zwiebach, 2002;
Coletti {\it {\it et al.}}, 2005).  These field redefinitions make it
very difficult to interpret simple physical properties of a system in
the variables natural to string field theory.  This is a generic
problem for background-independent theories, but some systematic way
of dealing with these different descriptions of physics needs to be
found for us to sensibly interpret and analyze multiple vacua within a
single formulation of string field theory.

Closely related to the issue of field redefinitions is the issue of
gauge fixing.  To perform explicit calculations in string field
theory, the infinite gauge symmetry (\ref{eq:SFT-gauge}) must be
fixed.  One standard approach to this is the ``Feynman-Siegel'' gauge,
where all states are taken to be annihilated by a certain ghost field.
For string fields near $\Psi = 0$ this is a good gauge fixing.  For
larger string fields, however, this gauge fixing is not valid (Ellwood
\& Taylor, 2001b).  Some string field configurations have no
representative in this gauge, and some have several (Gribov
ambiguities).  If for example one tries to continue the potential
graphed in Figure~\ref{f:potential} to negative $\varphi$ much below
the perturbative unstable vacuum or to positive $\varphi$ much past
the stable vacuum, the calculation cannot be done in Feynman-Siegel
gauge.  Currently no systematic way of globally fixing the gauge is
known.  This issue must be better understood to fully analyze the
space of vacua classically and to define the quantum theory.  For
example, it should be possible in principle to describe a two-D-brane
state in the Witten OSFT starting in the background with a single
D-brane.  This would correspond to a configuration satisfying the
equation of motion, but with energy above the perturbative vacuum by
the same amount as the stable vacuum is below it.  In this 2 D-brane
vacuum there would be 4 copies of each of the perturbative open string
states in the original model.  No state of this kind has yet been
found, and it seems likely that such a state cannot be identified
without a better approach to global gauge fixing.  It is interesting
to note that the analytic solution by Schnabl uses a different gauge
choice than Feynman-Siegel gauge; it will be interesting to see if
this gauge has better features with regard to some of the problems
mentioned here.

The open string field theory we have discussed here is a theory of
bosonic strings.  Attempting to quantize this theory is problematic
because of the bosonic closed string tachyon, which leads to
divergences and which is still poorly understood\footnote{Recent work
suggests, however, that even this tachyon may condense to a physically
sensible vacuum (Yang \& Zwiebach, 2005)}.  To discuss the quantum
theory we should shift attention to supersymmetric open string field
theory, which is tachyon free.  Witten's approach to describing OSFT
through a cubic action encounters problems for the superstring due to
technical issues with ``picture changing'' operators.  Although it may
be possible to resolve these issues in the context of Witten's cubic
formulation (Arefeva {\it et al.}, 1990), an approach which may be
more promising was taken by Berkovits (2001a, 2001b), where he
developed an alternative formulation of the open superstring field
theory.  This formulation is more like a Wess-Zumino-Witten model than
the Chern-Simons model on which (\ref{eq:SFT-action}) is based.  The
action has an infinite number of terms but can be written in closed
form.  Some analysis of this model using level truncation (see Ohmori,
2003 for a review) gives evidence that this framework can be used to
carry out a parallel analysis to that of the bosonic theory, and that
disconnected open superstring vacua can be described using this
approach, at least numerically.  At the classical level, the same
problems of field redefinition, lack of analytic tools, and gauge
fixing must be tackled.  But in principle this is a promising model to
extend to a quantum theory.  In principle, a complete quantum theory
of open strings must include closed strings, since closed strings
appear as intermediate states in open string scattering diagrams
(indeed in some sense this is how closed strings were first
discovered, as poles in open string scattering amplitudes).  It should
then in principle be possible to compute closed string scattering
amplitudes using OSFT.  A much more challenging problem, however, is
turning on nonperturbative deformations of closed string fields in the
open string language.  The simple version of this would be to deform a
modulus such as the dilaton by a constant value.  Much more
challenging would be to identify topologically distinct closed string
vacua as quantum states in a single OSFT.  Such a construction is well
beyond any tools currently available.  Since open string field theory
seems better understood than closed string field theory this is
perhaps a goal worth aiming at.  In the next section, however, we
describe the current state of direct constructions of closed string
field theory.

\section{Closed string field theory}

A direct formulation of closed string field theory is more complicated
than the theory for open strings.  In closed string field theory, the
string field $\Psi[X (\sigma)]$ has a field expansion
(\ref{eq:closed-expansion}) analogous to the open string field
expansion (\ref{eq:field-expansion}).  Writing an action for this
string field which reproduces the perturbative amplitudes of conformal
field theory is, however, much more complicated even in the bosonic
theory than the simple Witten action (\ref{eq:SFT-action}).

Using a generalization of the BRST formalism, Zwiebach (1992) developed a
systematic way of organizing the terms in a closed bosonic string
field theory action.  Unlike the Witten action, which has only
cubic interactions, Zwiebach's closed string field theory action
contains interaction terms at all orders.  The key to organizing this
action and making sure that it reproduces the standard closed string
perturbative expansion from CFT was finding a way of systematically
cutting apart Riemann surfaces (using ``Strebel differentials'') so
that each Riemann surface can be written in a unique way in terms of
propagators and vertices.  This approach is based very closely on the
geometry of the string world-sheet and it seems to give a
complete formulation of the bosonic theory, at least to the same
extent that Witten's theory describes the open bosonic string.

In closed string field theory there are massless fields corresponding
to marginal deformations of the closed string background.  Such
deformations include a modification of the string coupling, which is
encoded in the dilaton field $\phi (x)$ through $g = e^{\phi}$.  For
closed string field theory to be background independent, it needs to
be the case that turning on these marginal deformations can be
accomplished by simply turning on the fields in the SFT.  For example,
it must be the case that the string field theory defined with string
coupling $g$ has a background described by a certain field
configuration $\Psi'$, such that expanding the theory around this background
gives a theory equivalent to the SFT defined in a background with a
different string coupling $g'$.  This background independence was
shown for infinitesimal marginal deformations by Sen and Zwiebach 
(Sen \& Zwiebach, 1994a, 1994b).  This shows that closed string field
theory is indeed 
background independent.  It is more difficult, however,  to
describe a finite marginal deformation in the theory.  This problem is
analogous to the problem discussed in open string field theory of
describing a finite marginal deformation of the gauge field or
position of a D-brane, and there are similar technical obstacles to
resolving the problem.  This problem was studied for the dilaton and
other marginal directions by Yang and Zwiebach (2005a, 2005b).
Presumably similar techniques should resolve this type of marginal
deformation problem in both the open and closed cases.  A resolution
of this would make it possible, for example, to describe the moduli
space of a Calabi-Yau compactification using closed string field
theory.  One particularly interesting question is whether a
deformation of the dilaton to infinite string coupling, corresponding
to the M-theory limit, can be described by a finite string field
configuration; this would show that the background-independence of
string field theory includes M-theory.

To go beyond marginal deformations, however, and to identify, for
example, topologically distinct or otherwise disjoint vacua in the
theory is a much greater challenge.  Recently, however, progress has
been made in this direction also using closed string field theory.
Zwiebach's closed bosonic string field theory can be used to study the
decay of a closed string tachyon in a situation parallel to the open
string tachyon discussed in the previous section.  It has been shown
(Okawa \& Zwiebach, 2004a) that the first terms in the bosonic closed
string field theory give a nonperturbative description of certain
closed string tachyons in accord with physical expectations.  The
situation here is more subtle than in the case of the open string
tachyon, since the tachyon occurs at a point in space where special
``twisted'' modes are supported, and the tachyon lives in these
twisted modes, but as the tachyon condenses, the process affects
physics in the bulk of space time further and further from the initial
twisted modes.  This makes it impossible to  identify the new
stable vacuum in the same direct way as was done in OSFT, but the
results of this analysis do suggest that closed string field theory
correctly describes this nonperturbative process and should be capable
of describing disconnected vacua.  Again, however, presumably similar
complications of gauge choice, field redefinitions, and quantum
definition will need to be resolved to make progress in this
direction.

Because of the closed string bulk tachyon in the bosonic theory, which
is not yet known to condense in any natural way, the bosonic theory
may not be well-defined quantum mechanically.  Again, we must turn to
the supersymmetric theory.  Until recently, there was no complete
description of even a classical supersymmetric closed string field
theory.  The recent work of Okawa and Zwiebach (2004b) and of
Berkovits, Okawa and Zwiebach (2004), however, has led to an
apparently complete formulation of a classical string field theory for
the heterotic string.  This formulation combines the principles
underlying the construction by Berkovits of the open superstring field
theory with the moduli space decomposition developed by Zwiebach for
the bosonic closed string field theory.  Interestingly, for apparently
somewhat technical reasons, the approach used in constructing this
theory does not work in any natural way in the simpler type II theory.
The action of the heterotic superstring field theory has a
Wess-Zumino-Witten form, and contains an infinite number of
interactions at arbitrarily high orders.  The development of a SUSY
CSFT makes it plausible for the first time that we could use a
background-independent closed string field theory to address questions
of string backgrounds and cosmology.  Like the open bosonic theory
discussed in the previous section, this closed string field theory can
be defined in level truncation to give a well-defined set of
interaction terms for a finite number of fields, but it is not known
in any precise way what the allowed space of fields should be or how
to quantize the theory.  These are important problems for future work
in this area.

\section{Outlook}

We have reviewed here the current state of understanding of string field
theory and some recent developments in this area.  String field theory
is currently the only truly background-independent approach to string
theory.  We have reviewed some recent successes of this approach, in
which it was explicitly shown that distinct vacua of open string field
theory, corresponding to dramatically different string backgrounds,
appear as solutions of a single theory in terms of a single set of
degrees of freedom.  While much of the work concretely confirming this
picture in string field theory was numerical, it seems likely that
further work in the near future will provide a better analytic
framework for analyzing these vacua, and for understanding how open
string field theory can be more precisely defined, at least at the
classical level.

We described open string field theory in some detail, and briefly
reviewed the situation for closed string field theory.  While gravity
certainly requires closed strings, it is not yet clear whether we are
better off attempting to directly construct closed string field theory
by starting with the closed string fields in a fixed gravity
background, or, alternatively, starting with an open string field
theory and working with the closed strings which arise as quantum
excitations of this theory.  On the one hand, open string field theory
is better understood, and in principle includes all of closed string
physics in a complete quantum formulation.  But on the other hand,
closed string physics and the space of closed superstring vacua seems
much closer in spirit to closed string field theory.  Recent advances
in closed superstring field theory suggest that perhaps this is the
best direction to look in if we want to describe cosmology and the
space of closed string vacua using some background-independent
formulation of string theory along the lines of SFT.

We reviewed some concrete technical problems which need to be
addressed for string field theory, starting with the simpler OSFT, to make the
theory better defined and more useful as a tool for analyzing classes
of solutions.  Some problems, like gauge fixing and defining the space
of allowed states, seem like particular technical problems which come
from our current particular formulation of string field theory.  
Until we can solve these problems, we will not know for sure whether
SFT can describe the full range of string backgrounds, and if so how.
One
might hope that these problems will be resolved as we understand the
theory better and can find better formulations.
One hope may be that we might find a completely different approach
which leads to a complementary description of SFT.  For example, the
M(atrix) model of M-theory can be understood in two ways: first as a
quantum system of D0-branes on which strings moving in 10 dimensions
end, and second as a regularized theory of a quantum membrane moving
in 11 dimensions.  These two derivations give complementary
perspectives on the theory; one might hope for a similar alternative
approach which would lead to the same structure as SFT, perhaps even
starting from M-theory, which might help elucidate the mathematical
structure of the theory.

One of the problems we have discussed, however, seems generic to all
background-independent theories.  This is the problem of field
redefinitions.  In any background-independent theory which admits
numerous solutions corresponding to different perturbative
backgrounds, the natural degrees of freedom of each background will
tend to be different.  Thus, in any particular formulation of the
theory, it becomes extremely difficult to extract physics in any
background whose natural variables are different.  This problem is
already very difficult to deal with at the classical level.  Relating
the degrees of freedom of Witten's classical open string field theory
to the natural fields of conformal field theory in order to describe
familiar gauge physics, open string moduli, or the dynamical tachyon
condensation process makes it clear that simple physics can be
dramatically obscured by the choice of variables natural to string
field theory.  This problem becomes even more challenging when quantum
dynamics are included.  QCD is a simple example of this; the physical
degrees of freedom we see in mesons and baryons are very difficult to
describe precisely in terms of the natural degrees of freedom (the
quarks and gluons) in which the fundamental QCD Lagrangian is
naturally written.  Background independent quantum gravity seems to be
a similar problem, but orders of magnitude more difficult.

Any quantum theory of gravity which attempts to deal with the
landscape of string vacua by constructing different vacua as solutions
of a single theory in terms of a single set of degrees of freedom will
face this field-redefinition problem in the worst possible way.
Generally, the degrees of freedom of one vacuum (or
metastable vacuum) will be defined in terms of the degrees of freedom
natural to another vacuum (or metastable vacuum) through an extremely
complicated, generically quantum, field redefinition of this type.
This presents a huge obstacle to achieving a full understanding of
quantum cosmology.  This obstacle is very concrete in the case of
string field theory, where it will make it difficult to describe the
landscape of string vacua in  the language of a common theory.  It is
also, however a major obstacle for any other attempt to construct a
background-independent formulation of quantum gravity (such as loop
quantum gravity or other approaches reviewed in this book).  Only the
future will tell what the best means of grappling with this problem
may be, or if in fact this is the right problem to pose.  Perhaps
there is some radical insight not yet articulated which will  make it
clear that we are asking the wrong questions, or posing these
questions in the wrong way.

Two more fundamental issues which must be confronted if we wish to use
string field theory to describe cosmology are the issues of
observables and of boundary conditions and initial conditions.  These
are fundamental and unsolved issues in any framework in which we
attempt to describe quantum
physics in an asymptotically de Sitter or
metastable vacuum.  As yet there are no clear ways to resolve these
issues in SFT.  One interesting possibility, however, is that by
considering string field theory on a space-time with all spatial
directions compactified, these issues could be somewhat resolved.  In
particular, one could consider quantum OSFT on an unstable D-brane
(or a brane/antibrane pair for the supersymmetric OSFT or the closed
heterotic SFT without D-branes) on the background $T^9\times {\bf R}$.
The compactification provides an IR cutoff, and by putting in UV
cutoffs through level truncation and a momentum cutoff, the theory
could be approximated by a finite number of quantum mechanical degrees
of freedom.  This theory could be studied analytically, or, like
lattice QCD, one could imagine simulating this theory and getting some
approximation of cosmological dynamics.  If SFT is truly background
independent, quantum excitations of the closed strings should have
states corresponding to other compactification topologies, including
for example $T^3 \times X$ where $X$ is any flux compactification of
the theory on a Calabi-Yau or other 6D manifold.  Quantum fluctuations
should also allow the $T^3$ to contain inflating regions where the
energy of $X$ is positive, and one could even imagine eternal
inflation occurring in such a region, with bubbles of other vacua
branching off to populate the string landscape.  Or one could imagine
some other dynamics occurring, demonstrating that the landscape
picture is incorrect.  It is  impractical with our current
understanding to implement such a computation, and
presumably the detailed physics of any inflating region of the
universe would require a prohibitive number of degrees
of freedom to describe.  Nonetheless, if we can
sensibly quantize open superstring field theory, or a closed string
field theory, on $T^9$ or another completely compact space, it may in
some sense be the best-defined background independent formulation of
string theory in which to grapple with issues of cosmology.

\section*{Acknowledgements}

This work was supported in part by the DOE under contract
\#DE-FC02-94ER40818, and in part by the Stanford Institute for
Theoretical Physics (SITP).  The author would also like to thank
Harvard University for hospitality during part of this work.
Thanks to Barton Zwiebach for helpful discussions and comments on an
earlier version of this manuscript.

\end{document}